\documentclass[aip,reprint]{revtex4-1}
\usepackage{color}
\usepackage{ulem}
\usepackage{graphicx}

\begin{document}

\title{Enhancing light transmission through a random medium with inhomogeneous scattering and loss}

\author{Raktim Sarma}
\affiliation{Department of Applied Physics, Yale University, New Haven, CT, 06520, USA}
\author{Alexey Yamilov}
\email{yamilov@mst.edu}
\affiliation{Department of Physics, Missouri University of Science and Technology, Rolla, Missouri 65409,USA}
\author{Hui Cao}
\email{hui.cao@yale.edu}
\affiliation{Department of Applied Physics, Yale University, New Haven, CT, 06520, USA}

\begin{abstract}

We enhanced the total transmission of light through a disordered waveguide with spatially inhomogeneous scattering and loss by shaping the incident wavefront of a laser beam. Using an on-chip tapered lead, we were able to access all input modes in the waveguide with a spatial light modulator. The adaptive wavefront shaping resulted in selective coupling of input light to high transmission channels, which bypassed the regions of higher scattering and loss in the waveguide. The spatial inhomogeneity of scattering and loss led to redirecting of energy flux to optical paths with less scattering and loss to maximize total energy transported through the system. This work demonstrates the power of wavefront shaping in coherent control of light transport in inhomogeneous scattering media, which are common in real applications.

\end{abstract}

\maketitle

In recent years there have been rapid advances in coherent control of light propagation in strong scattering media \cite{MoskNatPhoReview}. It has been shown that light can be focused inside or through a turbid medium by shaping the input wavefront \cite{VellekoopReview}, which enables image transmission through an opaque material \cite{PopoffNatCom}. Wavefront shaping techniques have also been used to enhance the total transmission of light through a diffusive system via selective coupling of incident light to high transmission channels  \cite{MoskPRL,ChoiPRB,Kim12,YuPRL,Popoff2014,PRL_16}. These studies have important implications in biophotonics and biomedical applications \cite{ChoiReview, ParkReview}. However, in real samples such as biological tissues, the amount of light scattering often varies spatially. So far all the samples in wavefront shaping experiments are homogeneous, namely, the scattering strength is constant everywhere. Coherent control of light transport has not been demonstrated in inhomogeneous samples and the power of wavefront shaping in such systems is not known.

Light absorption is common in optical systems, and it can strongly modify high transmission channels. With strong absorption uniformly spread across a scattering medium, the diffusive transport of light in the maximum transmission channel turns into quasi-ballistic \cite{Seng14}.  In reality, optical absorbers are often distributed non-uniformly in random samples, and the high transmission channels redirect the energy flow to circumvent the absorbing regions to minimize attenuation \cite{Seng15}. These results are obtained from numerical simulations, and there has been no experimental study yet. Further, it is not clear what will happen when both scattering and absorption are spatially inhomogeneous.

In this Letter, we adopt the adaptive wavefront shaping approach to enhance light transmission through a disordered waveguide with spatially inhomogeneous scattering and loss. The silicon waveguide contains randomly distributed air holes within photonic crystal sidewalls. The degree of input control is much higher than that in the open slab geometry, thanks to an on-chip tapered lead. Light transport inside the two dimensional waveguide can be directly probed from the third dimension. After optimizing input wavefront to enhance the total transmission, we observe that optical waves bypass the region of higher scattering and loss in the waveguide. The spatial inhomogeneity of scattering and loss leads to redirecting of energy flux to optical paths with less scattering and loss, in order to maximize the total energy transported through the system. The experimental data agree to the numerical simulation results, revealing how a high transmission channel is modified by spatially inhomogeneous scattering and loss.

The disordered waveguide was fabricated in a silicon-on-insulator (SOI) wafer. The thickness of the silicon layer and of the buried oxide were 220 nm and 3 $\mu$m, respectively. The patterns were made by electron beam lithography and etched by an inductively-coupled-plasma (ICP) reactive-ion-etcher (RIE). Figure 1 is the scanning electron microscope (SEM) image of a fabricated sample. The waveguide is $L=60$ $\mu$m long and $W = 20$ $\mu$m wide. It contained a two-dimensional (2D) random array of air holes. While propagating in the waveguide, light is scattered both in plane and out of plane by the air holes. The out-of-plane scattering can be treated as loss, and the material absorption at the probe wavelength ($\lambda = 1510$ nm) is negligible \cite{Dz}.

To introduce spatially inhomogeneous scattering and loss, we varied the size and density of air holes in the waveguide. In a central region of diameter 10 $\mu$m, the air holes are larger and denser (hole diameter = 150 nm, air filling fraction = 15 $\%$), leading to stronger in-plane scattering and out-of-plane scattering. Outside this region the scattering and loss are weaker, as the air holes are smaller (diameter = 90 nm) and the filling fraction is lower (6 $\%$).

The relevant parameters to describe light propagation in the disordered waveguide are the transport mean free path $\ell$ and the diffusive dissipation length $\xi_a$. Their values in the two regions of different air hole size and density were extracted from the measurement of intensity distributions and fluctuations in two separate waveguides with homogeneous scattering and loss \cite{Sarma1_15}. In the central region, $\ell = 1$ $\mu$m and $\ell = 2.5$ $\mu$m; in the surrounding region, $\xi_a = 13$ $\mu$m and $\xi_a = 31$ $\mu$m.

The waveguide had reflecting sidewalls made of a triangular lattice of air holes (diameter = 360 nm, lattice constant = 550 nm). It supported an in-plane photonic bandgap at the probe wavelength, that confined the scattered light within the waveguide. The incident light was injected from the edge of the wafer to a silicon ridge waveguide. Due to the refractive index mismatch between silicon and air, the light could only excite the lower-order modes of the ridge waveguide, limiting the number of input modes that could be controlled by wavefront shaping. To increase the degree of input control, we designed and fabricated a tapered waveguide that served as a lead to the disordered waveguide \cite{PRL_16}. The tapering angle was $15^{\circ}$, and the waveguide width was reduced from 330 $\mu$m to $20$ $\mu$m over a length of 578 $\mu$m.  The wider waveguide at the input supported many more lower-order modes that were converted to higher-order modes by the taper. The numerical simulation confirmed that the number of waveguide modes excited at the air-silicon interface by the incident light is significantly larger than the number of transmission channels in the disordered waveguide $N = 75$.

\begin{figure}
\centering{\includegraphics[width=3.5in]{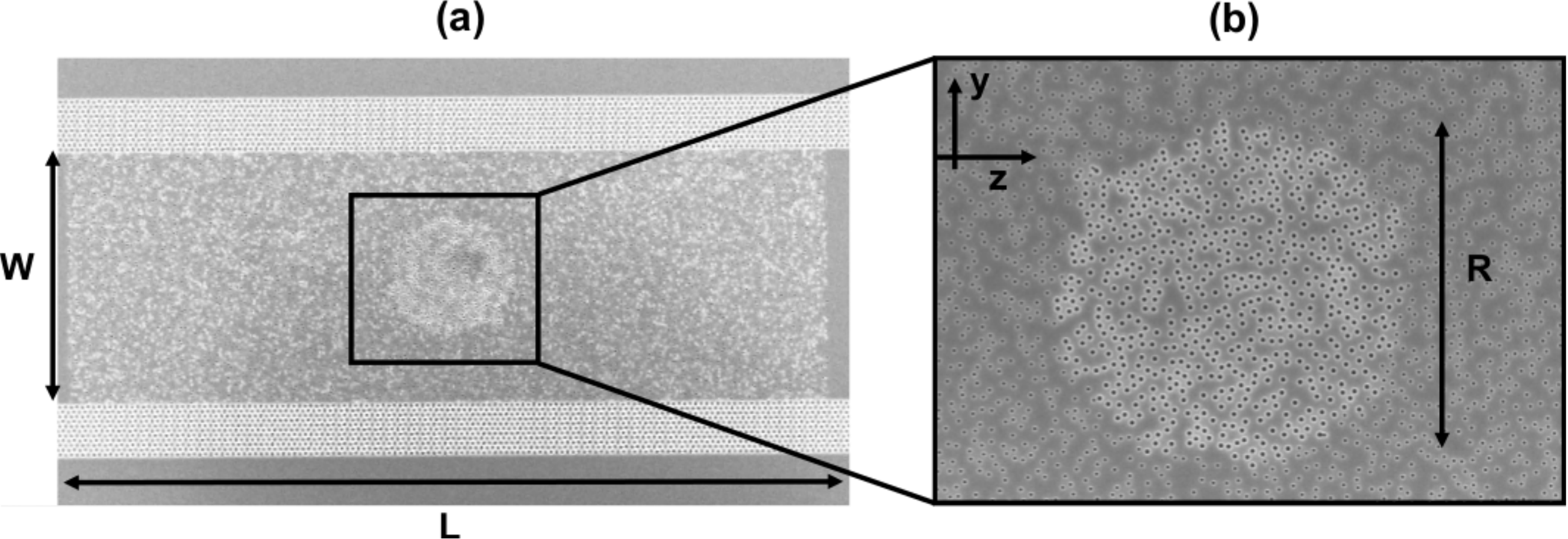}}
\caption{
\label{fig:sem}
2D disordered waveguide with inhomogeneous scattering and loss. (a) Top-view scanning electron micrograph (SEM) of the fabricated silicon waveguide that consists of randomly positioned air holes. The waveguide width $W = 20$ $\mu$m, and length $L = 60$ $\mu$m. A circular region of diameter $ 10$ $\mu$m at the center of the waveguide has larger and denser air holes (hole diameter = 150 nm, the air filling fraction = 15 $\%$). Outside this region, the air holes are smaller (diameter = 90 nm) and the filling fraction is lower (6 $\%$). The sidewalls of the waveguide are made of a triangular lattice of air holes (diameter = 360 nm, lattice constant = 505 nm), which supports an in-plane photonic bandgap at the wavelength $\lambda = 1.51$ $\mu$m. (b) Magnified SEM of the central region of the disordered waveguide showing air holes of two different sizes and densities.}
\end{figure}

\begin{figure}
\centering{\includegraphics[width=3.5in]{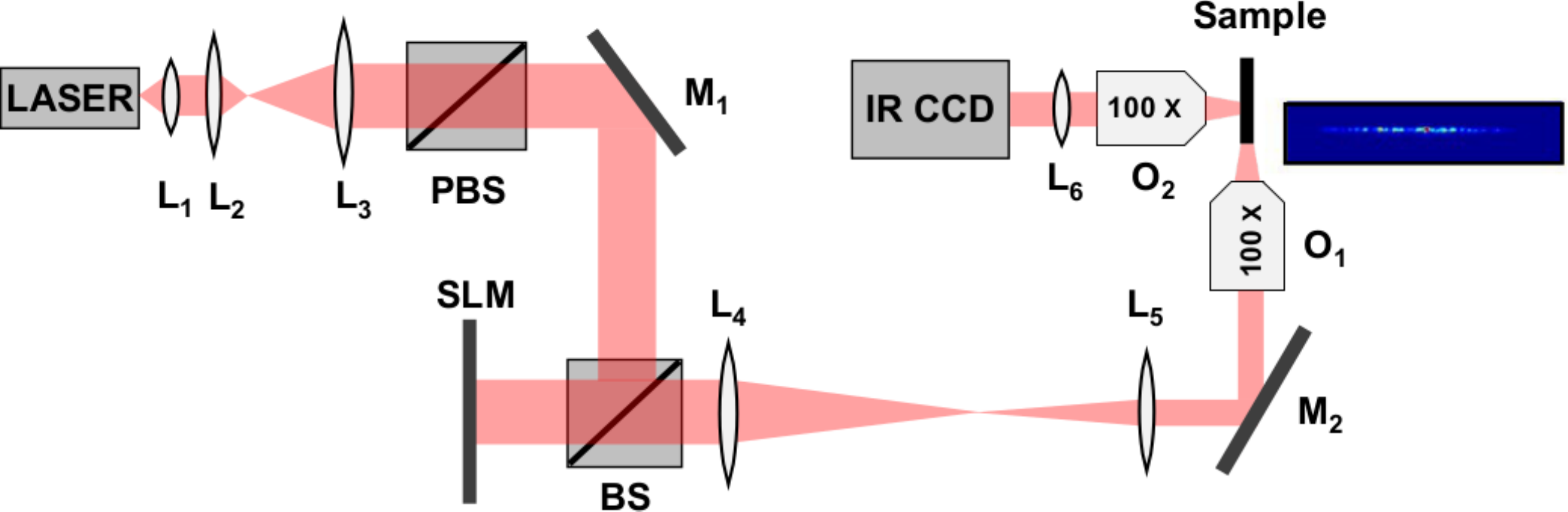}}
\caption{
\label{fig:experiment}
A schematic of the wavefront shaping experiment setup. A laser beam (HP 8168F) at $\lambda=1510$ nm is collimated (by lens $L_1$), expanded (by $L_2$, $L_3$), and linearly polarized (by a polarized beam splitter PBS) before being modulated by a phase-only SLM (Hamamatsu X10468). Two lens ($L_4$, $L_5$) are used to project the SLM plane to the pupil plane of an objective $O_1$ ($100 \times$, NA = 0.7), and the edge of the coupling waveguide is placed at the focal plane of the objective. The light scattered out of the sample plane is collected by a second objective $O_2$ ($100 \times$, NA = 0.7) and imaged to an InGaAs camera (Xenics XEVA 1.7-320) by a tube lens ($L_6$). $M_1$ and $M_2$ are mirrors, BS is an unpolarized beam splitter. The inset is an optical image of the illumination line on the front facet of the coupling waveguide, created by modulating the phase of the SLM pixels. }
\end{figure}

To control light transport in the disordered waveguide, we adopted the adaptive wavefront shaping scheme that we had recently implemented for 2D on-chip waveguides \cite{PRL_16}. The setup is shown schematically in Fig. 2. A monochromatic laser beam was collimated, expanded and linearly polarized. It is then phase modulated by a spatial light modulator (SLM). The SLM plane was demagnified and projected to the pupil plane of an objective. At the focal plane of objective lied the front facet of the coupling waveguide. We imposed one-dimensional phase modulation on the SLM to create a line of illumination for the coupling waveguide, as shown in the inset of Fig. 2.  To map the spatial distribution of light intensity, $I(y,z)$, inside the disordered structure, the out-of-plane scattered light was collected by a second objective and projected to an InGaAs camera.

To enhance the total transmission through the disordered waveguide, we chose the feedback-based optimization technique, which was robust against measurement noise. The cost function $S$ was given by the ratio of the cross-section integrated intensity of light at the back end of the waveguide to that at the front end. We used the continuous sequential algorithm to maximize $S$ by adjusting the phase of SLM pixels \cite{VellekoopReview}. Figures 3(a) and (b) show the intensity distribution $I(y,z)$ for an unoptimized input and an optimized input, respectively. When the input wavefront was not optimized, the light intensity decreased with the depth in the disordered waveguide. Stronger out-of-plane scattering brightened the central region that had larger and denser air holes. In contrast, the optimized input wavefront made the central region dark, meanwhile the intensities on both sides of this region increased. Such changes indicated that light bypassed the central region with higher scattering and loss to maximize the total energy transported through the medium.

For a better understanding of the experimental results, in order to understand the energy flow for optimized input, we performed a numerical simulation \cite{2014_Groth_Kwant} to calculate the ensemble averaged Poynting vector $\vec{J}(y,z)$ for an optimized input of a 2D disordered waveguide with all parameters equal to the experimental values. The continuous sequential algorithm  was used to optimize the total transmission via phase-only modulation of the input wavefront. The total transmission was increased from $3.2\%$ with unoptimized input to $42\%$ with optimized input. Figure 4(a) plots the magnitude and direction of $\vec{J}(y,z)$ in the disordered waveguide for an optimized input. The optimized input wavefront made the energy flux circumvent the central region with higher scattering and loss, in agreement to the experimental result. Further, Fig. 4(b) shows the magnitude and direction of $\vec{J}(y,z)$ of the maximum transmission channel, which resembles that of the optimized input in Fig. 4(a). This result suggested that with the optimized input wavefront, light transport was dominated by the maximum transmission channel. This was confirmed by decomposing the optimized input wavefront by the transmission eigenchannels. The contribution from the maximum transmission channel was significantly larger than all other channels. Therefore, the optimization of incident wavefront led to selective coupling of light to the high transmission channels.

\begin{figure}
\centering{\includegraphics[width=2.85in]{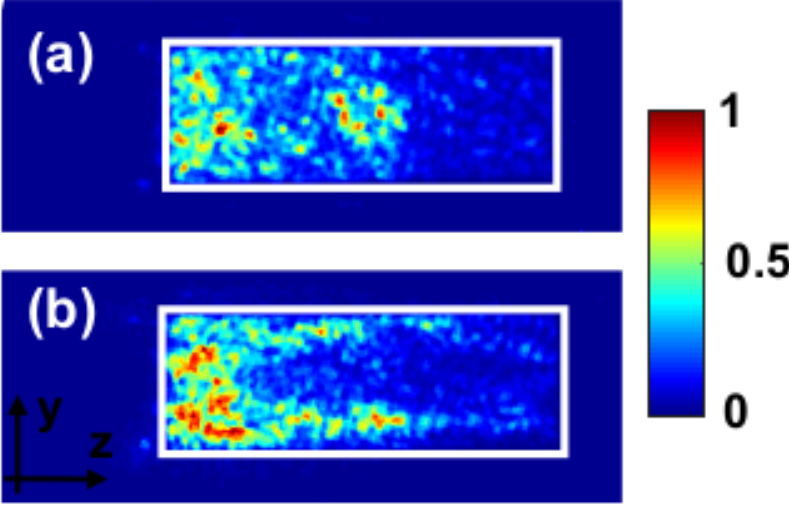}}
\caption{\label{fig:Results}  Optimizing the incident wavefront to enhance light transmission through the disordered waveguide with spatially inhomogeneous scattering and loss. Experimentally measured 2D intensity distribution $I(y,z)$ inside the waveguide shown in Fig. 1 for (a) unoptimized input fields, (b) optimized input for maximum cost function $S$. The white box marks the boundary of the disordered waveguide.}
\end{figure}

\begin{figure}
\centering{\includegraphics[width=3.5in]{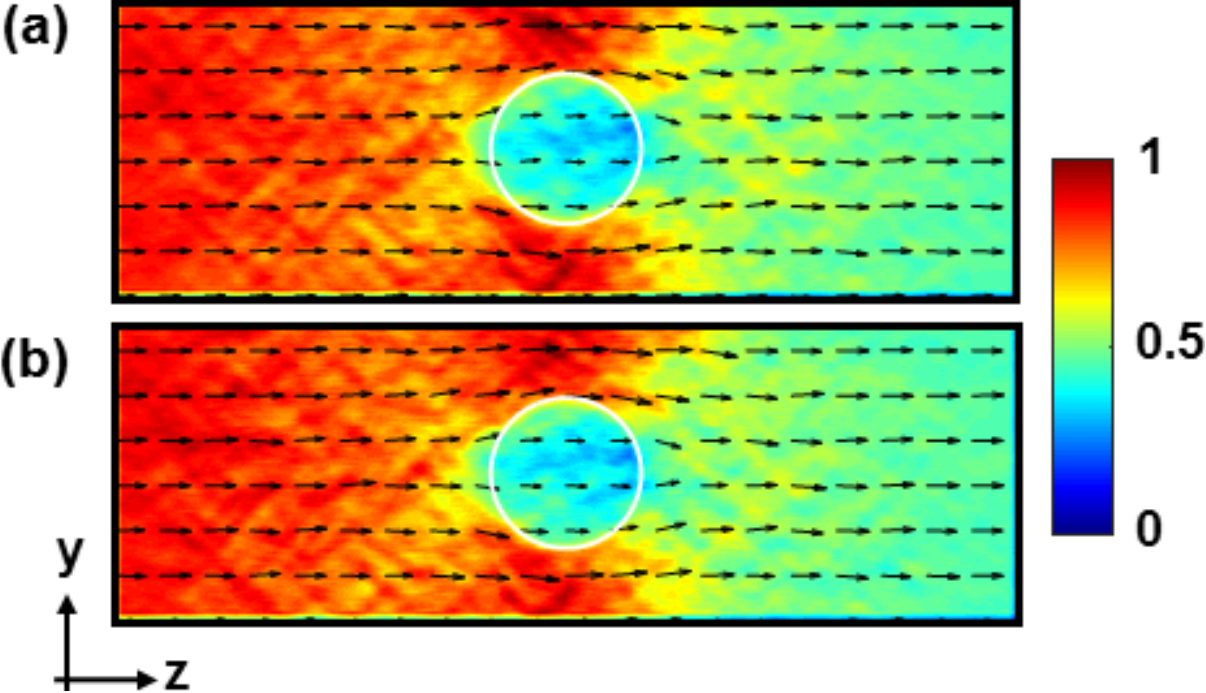}}
\caption{
\label{fig:simulations}
Numerical simulation of the ensemble averaged Poynting vector $\vec{J}(y,z)$ of light inside the 2D disordered waveguide with spatially inhomogeneous scattering and loss. The magnitude of $\vec{J}(y,z)$ is shown by color plot, and its direction is shown by the arrows. The input field in (a) is optimized to maximize total transmission. With optimized input wavefront, the optical waves bypass the region of higher scattering and loss in the middle of the waveguide (denoted by a while circle). (b) shows $\vec{J}(y,z)$ for the maximum transmission channel, which is nearly identical to that in (a), indicating the optimized input field couples mostly to the maximum
transmission channel.
}
\end{figure}

In summary, we enhanced light transmission through a 2D waveguide with spatially inhomogeneous scattering and loss by shaping the wavefront of incident light. Using a tapered lead, we were able to access all input modes by a spatial light modulator. The optimized wavefront selectively coupled light to high transmission channels, which bypass the regions of higher scattering and loss. This work demonstrated the power of wavefront shaping in controlling light transport in inhomogeneous scattering samples, which are common in real applications. In addition, our results may trigger further studies of on-chip disordered photonic nanostructures with spatially varying scattering strength and loss to mold the flow of light \cite{YamilovOL16}.

\begin{acknowledgments}

We acknowledge Chia-Wei Hsu, Douglas Stone, Hasan Yilmaz, and Seng Fatt Liew for useful discussions. This work was supported by the Office of Naval Research (ONR) under grant no. MURI N00014-13-0649, by the US-Israel Binational Science Foundation (BSF) under grant no. 2015509 and by the National Science Foundation (NSF) under grant no. NSF DMR-1205223. Facilities use was supported by YINQE and NSF MRSEC DMR-1119826.

\end{acknowledgments}

\end{document}